# Controlling spin pumping into superconducting Nb by proximity-induced spin-triplet Cooper pairs

A. K. Chan[1,2*], M. Cubukcu[2], X. Montiel[3], S. Komori[3], A. Vanstone[1,2], J. E. Thompson[3], G. K. Perkins[1], M. Blamire[3], J. Robinson[3], M. Eschrig [4], H. Kurebayashi[2] and L. F Cohen[1*]

[1]Department of Physics, Blackett Laboratory, Imperial College London, Prince Consort Road, London SW7 2AZ, UK

[2]Department of Electronic and Electrical Engineering and London Centre for Nanotechnology, 17-19 Gordon Street, London  WC1H 0AH, UK

[3]Department of Materials Science & Metallurgy, University of Cambridge, 27 Charles Babbage Road, Cambridge CB3 0FS

[4]Department of Physics, University of Greifswald, Wissen Lockt. Seit 1456, Germany

**Abstract**

**Proximity-induced long-range spin-triplet supercurrents, important for the field of superconducting spintronics, are generated in superconducting/ferromagnetic heterostructures when interfacial magnetic inhomogeneities responsible for spin mixing and spin flip scattering are present. The multilayer stack Nb/Cr/Fe/Cr/Nb has been shown to support such exotic currents when fabricated into Josephson junction devices. However, creating pure spin currents controllably in superconductors outside of the Josephson junction architecture is a bottleneck to progress. Recently, ferromagnetic resonance was proposed as a possible direction, the signature of pure supercurrent creation being an enhancement of the Gilbert damping below the superconducting critical temperature, but the necessary conditions are still poorly established. Consistent with theoretical prediction, we demonstrate conclusively that pumping pure spin currents into a superconductor is only possible when conditions supporting proximity-induced spin-triplet effects are satisfied. Our study is an important step forward for superconducting pure spin current creation and manipulation, considerably advancing the field of superconducting spintronics.**

**Introduction**

Superconductor (SC)/ ferromagnet (FM) interfaces are of great interest as potential candidates to exploit the spin degree of freedom in superconducting phenomena, leading to potential applications for cryogenic memory and novel computing technologies [1, 2]. As its cornerstone observation, the presence of a spatially varying magnetization at a SC/FM interface has been found to generate long range spin-polarized triplet supercurrents in the FM via a superconducting proximity effect in combination with spin mixing and spin flip scattering processes [1, 3, 4, 5, 6, 7, 8, 9, 10]. These proximity-induced triplet supercurrents are attractive for emerging applications in superconducting spintronics as transmitters of spin information within logic circuits, without incurring ohmic dissipation [1, 3, 4, 5, 6, 7, 11].

Creation of pure spin currents within a superconductor would also accelerate developments in the field, particularly if their generation was independent of the Josephson junction architecture. Houzet [12] proposed such a route by using ferromagnetic resonance (FMR). It has been long established that spin pumping is a direct means to generate and transport pure spin currents from a FM into adjacent materials using the magnetization dynamics [*M(t)*] associated with ferromagnetic resonance [13, 14, 15]. The generation of spin currents from the FM affects the magnetization dynamics via the enhancement of the effective Gilbert damping $\alpha$.

Using FMR, it has been demonstrated [16, 17] in FM/SC systems with clean interfaces that the SC blocks the transport of a dynamically driven spin current. The creation of equilibrium (spin-zero) spin-singlet Cooper pairs depletes the number of normal electrons in the SC, and therefore supresses Andreev reflection at the FM/SC interface. Below $T_c$, singlet blocking dominates and results in a decrease of $\alpha$ [18, 19].

However, recently Jeon *et al.* [20] demonstrated that pure spin currents appeared to survive within superconducting Nb when spin pumping by FMR into Pt/Nb/Py/Nb/Pt stacks, but only when Pt a strong spin-orbit coupling (SOC) layer was included [20, 21, 22, 23]. The identifying feature was a sustained broadening of the FMR linewidth associated with an increase in $\alpha$, as the sample was cooled below $T_c$. These observations are consistent with the theoretical predictions that SOC in combination with the ferromagnetic exchange interaction can also generate conditions for the formation of spin triplet superconductivity [24, 25]. Although an important and encouraging development, the influence of the spin triplet channel on the ability to pump pure spin currents into a superconductor was obscured by the non-local character of the Fe and Pt interaction, indeed the source of much subsequent discussion [18, 19, 26, 27]. Therefore, the precise role of the spin triplet channel on pure spin pumping into a superconductor remains an outstanding question.

To explore this research question explicitly, we study a system known to support long range proximity-induced spin triplets. Josephson junctions made up of Nb/Cr/Fe/Cr/Nb layers have been recently shown to carry supercurrents through a significant thickness of Fe due to inhomogeneous magnetism at the Fe/Cr interfaces [28, 29]. Therefore, in this study we examine the FMR properties of Nb/Fe/Nb and Nb/Cr/Fe/Cr/Nb thin-film multilayer stacks in the normal and superconducting state, as shown schematically in Figure 1. Without Cr present the conditions to form proximity-induced spin triplet states within the superconductor are not satisfied. Moreover, in these samples the Fe layer shows in-plane uniaxial magnetic anisotropy, due to the sputtering growth conditions onto $SiO_2$/Si(001) substrates (as previously reported [30]), which places a further condition on the spin triplet state survival in finite magnetic fields necessary for FMR resonance. We demonstrate experimentally that when the applied magnetic field lies along the magnetic hard axis, an enhancement of the Gilbert damping persists to low temperatures in NbCr/Fe/Cr/Nb structures. However, when the field lies along the easy axis we do not observe this signature. We also show that the Gilbert damping enhancement is absent in Nb/Fe/Nb samples.

**Results**

We focus on samples grown with a 6 nm thick layer of Fe and a 32 nm thick layer of Nb, referred to as "*32Cr*" including a 1 nm thick layer of Cr, and "*32NCr*" when Cr is absent. A summary of these structures, and those shown in the SI, is given in Table 1.

**Magnetic characterization and anisotropy**

Although the Fe is polycrystalline, it shows in-plane uniaxial magnetic anisotropy. Figure 2(a) shows the magnetization *M(H)* data taken on sample 32Cr in different in-plane magnetic field orientations showing uniaxial magnetic easy and hard characteristics. This is also confirmed by our FMR measurements. Typical FMR spectra measured along the two directions, shown in Figure 2(b), display a significant difference in FMR resonance field $\mu_0 H_{res}$. The temperature dependence of $\mu_0 H_{res}$ for samples 32Cr and 32NCr are shown in Figure 2(c). The difference in resonance field $\mu_0 \Delta H_{res}$ between applying the applied dc field along the easy and hard directions, is shown in Figure 2(d). $H_{res}$ is higher in the Cr containing film. This behaviour is confirmed in a second series of samples 31Cr and 31NCr (see SI3).

The magnetic anisotropy $H_k$ can be extracted from the modified Kittel formula fitted to the FMR data, taken with the dc field applied along the easy and hard magnetic directions [31], as set out in equation 1:

$$f^2 = \left(\frac{\gamma \mu_0}{2\pi}\right)^2 [\{M_{\text{eff}} + H_{\text{res}} + H_k \sin^2(\phi - \phi_0) + H_{\text{shift}}^{\text{SC}}\} \\ \times \{H_{\text{res}} - H_k \cos 2(\phi - \phi_0) + H_{\text{shift}}^{\text{SC}}\}], \qquad (1)$$

where $\gamma$ is the gyromagnetic ratio, $\mu_0$ is the permeability of free space, $M_{\text{eff}}$ is the effective saturation magnetisation, $\phi$ is the angle of the film's magnetisation with respect to the easy axis magnetic direction, $\phi_0$ is the angle of the hard axis direction with respect to the easy axis direction, and $H_{\text{shift}}^{\text{SC}}$ is the contribution to the local field acquired in the superconducting state [32] (see SI5 for further details). Due to the relatively small values and sensitivity of $H_k$ compared to $H_{\text{res}}$ and $M_{\text{eff}}$, we simultaneously fit data taken with the applied magnetic field aligned along the easy and hard axis direction in order to extract single values for $H_k$ and $M_{eff}$. An example fit for 32Cr is shown in Figure 3(a). The anisotropy energy constant $K$ can then be calculated by the relation $K = H_k M_s/2$ and summarised in Table 1. We can observe fair agreement between $K$ extracted from two independent techniques, i.e., $M(H)$ and FMR. The size of the magnetic anisotropy is comparable to that of previous studies [30, 33].

**Gilbert damping**

The Gilbert damping $\alpha$ and the inhomogeneous broadening $\Delta H_0$ can be extracted from $\Delta H$ vs $f$ plots using the following equation [34]:

$$\Delta H = \Delta H_0 + \frac{4\pi\alpha}{\gamma}f. \qquad (2)$$

and an example fit is shown in Figure 3(b) for 32Cr. The presence of Cr at the interfaces increases $\mu_0 \Delta H_0$ by a factor of two to three (for Nb/Fe/Nb samples $\mu_0 \Delta H_0 \approx$ 10 mT and Nb/Cr/Fe/Cr/Nb samples $\mu_0 \Delta H_0$ lies between 20 mT to 30 mT) as expected for the increased interfacial magnetic disorder that occurs due to the Cr inter-diffusion at the Cr/Fe interface. Figure 3(c) shows $\alpha$ for sample *32Cr* as a function of temperature extracted from $\Delta H$, for microwave frequencies from $f$ = 12 GHz to $f$ = 20 GHz. Note that the $\alpha$ values shown are averaged over several measurements. We also checked the reproducibility of measurements by performing experiments with the microwave stripline placed in different regions of the sample (see supplementary SI6) – the error bars reflect the variability in measurement. In-phase ac susceptibility $\chi'_{ac}$(red) measurement of the bulk $T_c$ are plotted on the same graph. The onset of superconductivity is indicated by a decrease in $\chi'_{ac}$, with the gap fully opened when $\chi'_{ac}$ reaches a constant minimum value (see supplementary and figure SI2 for further information). For all frequencies, $\alpha$ is almost *T*-independent between 16 K and $T_c$. As the temperature is reduced below $T_c$, a significant temperature-dependent enhancement of $\alpha$ occurs for the field aligned along the hard-axis that is in stark difference from spin transport blocking caused by singlet Cooper pairs [16, 17, 18, 26, 27]. This clear enhancement of $\alpha$ with Cr samples below $T_c$ is indicative of higher rates of spin relaxation in the Fe layer when the Nb layers are superconducting [13, 14]. Since a spin current carries spin-angular momentum, this damping enhancement that increases as the sample is cooled suggests an increasingly strong spin-current transmission and dissipation as the superconducting gap opens, consistent with previous observations [20, 21, 22, 23]. When the same experiment is repeated for the field aligned along the magnetic easy axis, the monotonic enhancement of $\alpha$ is absent, as shown in Figure 3(c). Samples without Cr replicate this latter behaviour below $T_c$ regardless of the orientation of the magnetic field alignment with respect to the easy and hard magnetic axis. These observations are repeated in the second series of samples *31Cr* and *31NCr* (see supplementary Figure S6) and the results are consistent with those reported here.

We can compare our findings with theoretical prediction that resembles our experimental layout. In order to do this, we model the Fe/Cr/Nb systems by a F1/F2/S trilayer with a ferromagnetic layer F1, a ferromagnetic layer F2 (which has its spin magnetic moment misaligned with respect to F1 at a fixed angle) and a superconducting layer S. We treat only the simplest model, which in contrast to the experiment assumes that F2 precesses together with F1, allowing for an efficient numerical modelling of the space-time dependence following the treatment of Houzet [12]. We assume that the precession in the F2 layer occurs at a relative angle compared to than in the F1 layer. This angle difference can be explained by different interaction and crystalline structure in both F1 and F2 layers. In the theoretical calculation, the misalignment of spin at the Fe/Cr/Nb interface is modelled by a single unique misalignment parameter $\theta$. The theory takes into account the transparency of the interface ($t$), the strength of the exchange field ($J$) and the suppression of the superconducting gap energy at the interface ($\Delta$). It does not take into account the quasiparticle temperature dependence (so it does not capture the coherence peak feature), interface roughness, or the dependence of the spin current on momentum direction. A full description of this theoretical approach is given in [35]. Figure 3(e) sets out the theoretical prediction of $\alpha$ as a function of $\theta$ for optimum values of the parameters of $t$, $J$, $\Delta$ (i.e., parameters that give the maximum $S$ value in the full treatment). The Gilbert damping is normalized here by the value of the Gilbert damping above $T_c$ with $\theta = 0$. In the normal state $\alpha/\alpha_{T>Tc}$ increases as $\theta$ increases, indicative of an additional loss channel when the interfacial spins are misaligned to the precessing ferromagnet F1. This scenario can be mapped across to the experimental observation of 'easy' and 'hard' magnetic directions that we observe in the FMR resonance field.

**Discussion**

Let us consider the various temperature regimes that delineate the key experimental observations. Firstly, above $T_c$, we find that Cr containing samples have larger $\Delta H_0$ and larger Gilbert damping than the samples without Cr implying that the presence of the Cr always adds an additional spin relaxation channel. Although rather temperature insensitive above $T_c$, this channel presents a more efficient loss mechanism when the magnetic field is in the hard axis. Together this suggests that Cr creates a strong spin relaxation channel most likely related to a spin misorientation angle between an interfacial disordered Cr/Fe diffused magnetic layer and the spin current polarization [13, 36].

At $T_c$, the onset of superconductivity, in both the non-Cr containing samples and the Cr containing samples with the field along the easy axis, a peak is observed in $\alpha$. This is associated with the formation of the quasiparticle coherence peak [18, 27, 37]. The injected spin current (and therefore the damping) first increases due to the enhanced availability of quasiparticle states just below $T_c$, and then decreases upon further cooling as the quasiparticle states freeze out and the superconducting gap begins to fully open [17].

Below $T_c$, $\alpha(T)$ behaves in two different ways depending on the presence of Cr and the field orientation. Regardless of field orientation, samples without Cr show a decrease in $\alpha$. Similarly, regardless of the Cr layer, measurements with the field applied along the easy axis show a decrease in $\alpha$. Both scenarios indicate partial spin blocking below $T_c$ due to the onset of superconductivity.

In samples containing Cr when the field is along the hard axis, $\alpha$ increases steadily as the temperature is reduced well below $T_c$. The ac susceptibility curve marks where the transition to the superconducting state completes. This is a striking difference to the above scenario and the observation cannot be attributed to the quasiparticle population [18, 19]. The clear enhancement of $\alpha$ with Cr samples well below $T_c$ is indicative of higher rates of spin relaxation when the Nb layers are superconducting [13, 14]. Since a spin current carries spin-angular momentum, this damping enhancement suggests an increasingly strong spin-current transmission and dissipation as the superconducting gap opens, indicative of a spin triplet channel in the Nb. This is consistent with, although significantly larger than, previous observations [20, 21, 22, 23].

It is already established from Josephson junction measurements that below $T_c$, the spin misorientation between Fe and the inter-diffused Fe/Cr layer magnetisation produces long-range triplet correlations [12, 38, 39, 40]. Here we suppose that when the field is aligned along the magnetically hard axis the misorientation is sustained up to the dc fields required under the resonance condition to detect a triplet channel. Although the volume magnetization suggests macroscopic saturation at fields in the tens of mT (Figure S1), evidence from magnetotransport of Cr/Fe multilayers suggests canting of spins at this interface that survive to high magnetic field [28, 41]. The presence of equal-spin Cooper pairs provides a new channel to carry spin within the superconducting Nb where it dissipates due to the inherent spin-orbit coupling in that material. Sample series *31Cr* and *31NCr,* where the uniaxial magnetic anisotropy is measurably reduced (refer to Table 1 and SI Figures S1(b) and (c)), show a reduced upturn of $\alpha(T)$ along the hard axis (shown in SI6). This supports the statement that the introduction of Cr at the interface does not in itself provide a sufficient condition for the spin triplet channel unless the spin misalignment angle between Fe and the interfacial layer can be maintained at the FMR resonance field condition.

To discuss this more quantitatively, we discuss the damping enhancement by introducing parameter *S*, given by

$$S = \Delta\alpha/\bar{\alpha}(T > T_c), \tag{3}$$

where $\bar{\alpha}(T > T_c)$ s the average $\alpha$ value for $T > T_c$, and $\Delta\alpha$ is the difference between the average $\alpha$ values for $T < 0.8T_c$ and $T > T_c$, such that $\Delta\alpha = \bar{\alpha}(T < 0.8T_c) - \bar{\alpha}(T > T_c)$. When the dc magnetic field is applied along the hard axis of our Nb/Cr/Fe/Cr/Nb multilayer stack the spin channel strength created below the superconducting critical temperature is $S = 0.36$ for sample *32Cr* and $S = 0.14$ for sample *31Cr* (in the SI). This can be compared to the previous example of Pt/Nb/Py/Nb/Pt [20] where the spin-orbit coupling and ferromagnetic exchange were spatially separated, giving $S = 0.066$. The higher $S$ values produced by spin pumping into Nb/Cr/Fe/Cr/Nb multilayer stacks suggest that the spin flip and spin scattering processes provided by the single inhomogeneous interface allows a more efficient means of producing spin-triplet supercurrents in the superconductor. Furthermore, sample *32Cr* has a higher $S$ value than *31Cr*, indicating the higher magnetic anisotropy is crucial in generating the spin-triplet channel.

At the lowest temperatures, although the theory prediction and the experimental observation are in qualitative agreement there is a discrepancy in the magnitude of the effect by a factor of 6 in S parameter defined in equation 3, $S = 0.06$ theoretically and 0.35 experimentally. The most significant aspect that the theory does not consider is the fact that in the experiment the second ferromagnetic layer has a static magnetization, which in combination with the precessing magnetization in the first ferromagnet leads to a time-dependent Gilbert damping. The additional production of spin waves in the F2 layer likely leads to an underestimation of the Gilbert damping due to the fact that spins are pumped back from F2 to F1, a process that is absent in the experimental setup. A full calculation would require a numerical study of space-time-dependent solutions which is left for future work. Further aspects that could be improved on are taking into account a dependence of the spin current on momentum direction, and taking into account interface roughness and crystallinity. It is possible therefore that the experimental situation with static magnetization in F2, polycrystalline films, a finite degree of interfacial roughness and a range of misalignment angles optimises the strength of the spin triplet current beyond the scope of the current theoretical predictions. The parallels between theoretical and experimental behaviour are encouraging as they suggest that strong superconducting spin channels can be created under realistic experimental conditions.

To summarise our work, we have systematically investigated spin pumping in Nb/Cr/Fe/Cr/Nb and Nb/Fe/Nb multilayers and compared to theoretical prediction, with the specific goal to establish the impact of superconducting spin triplets on the spin pumping behaviour. We confirm that under optimum conditions for the existence of proximity-induced superconducting spin triplets, the signature for pure spin pumping into the superconducting Nb, namely a sustained $\alpha$ enhancement below the superconducting critical temperature, is manifest. These results conclusively show that superconducting pure spin currents can be created within a superconductor under a specific and well-defined set of interfacial conditions. The qualitative consistency between theory and experiment highlights the physical scenario and points to future directions to explore the influence of interfacial roughness and crystallinity. We also highlight that the FMR measurement method itself is a sensitive measure of magnetic relaxation, specifically in relation to spin current emission and dissipation, and that the damping enhancement parameter $S$ can be used as a relative measure of the strength of spin triplet current density between different types of heterostructure stacks. Overall, the work accelerates the possibility of utilising spin dependent supercurrents in low loss spintronic applications.

**Methods**

**Sample preparation.**

The heterostructures were grown on 5 mm × 5 mm $SiO_2$/Si(001) substrates by d.c. magnetron sputtering in an ultrahigh-vacuum chamber (base pressure of around $10^{-8}$ Pa) with a growth pressure of 1.5 Pa of Ar. Multiple substrates were placed on a rotating circular table that passed in series under stationary magnetrons, so that eight samples with different layer thicknesses could be grown in the same deposition run. This ensures that the interface properties of the samples presented closely match. The thickness of each layer was controlled by adjusting the angular speed of the rotating table at which the substrates moved under the respective magnetron target and the sputtering power. The thicknesses of Fe and Cr layers were kept constant at 6 nm and 1 nm respectively, while the thickness of the Nb layer varied at 31 and 32 nm.

**Superconducting transition measurement.**

A.c. electrical transport measurements were conducted with a four-point current–voltage method. The resistance $R$ (of a sample) versus temperature $T$ curves were obtained while decreasing $T$ (see supplementary information). From the $T$ derivative of $R$, $dR/dT$, the superconducting transition temperature $T_c$ was denoted as the $T$ value that exhibits the maximum of dR/dT.

In-phase a.c. susceptibility measurements were also performed to further confirm $T_c$ when externally dc fields are applied at typical FMR field strengths. Samples are placed within pick up coils with an in-plane drive and external dc fields can be applied in-plane. See SI2 for further information.

**Ferromagnetic resonance.**

Ferromagnetic resonance (FMR) is performed using a broadband coplanar waveguide (CPW) and ac-field modulation technique. The samples are placed face down on top of the CPWs where an insulator tape is used for electrical insulation. The magnetic field is applied along in-plane direction of the samples. The FMR absorption was measured for different frequencies typically ranging from 12 to 20 GHz from room temperature to low temperature at 2K. For each scan, the resonance field ($H_{res}$) and the half width at half maximum linewidth ($\Delta H$) of the FMR signal are determined by a fit using the field derivative of Lorentzian function (see SI3-5).

**Theory**

We calculate non-equilibrium spin current in a F1/F2/S trilayer where F1 is a precessing ferromagnetic, F2 is a misaligned ferromagnetic layer with misalignment angle θ, and S is a superconductor of thickness $d_S$. The misalignment between the magnetizations of the F1 and F2 layers induces equal-spin Cooper pairs across the entire trilayer. We consider a spin precession occurring in the x-y plane while the magnetization is tilted in the y-z plane. The trilayer is stacked along the x axis and we assume that the layers extend infinitely in the y-z plane. In the following, we focus on the stationary regime where both F1 and F2 layer magnetizations exhibit the time-dependency [12]:

$$J_{Fi}(t) = |J_{Fi}|(\sin\theta_{Fi}\sin\Omega t, \sin\theta_{Fi}\cos\Omega t, \cos\theta_{Fi}) \quad (4)$$

Where i=1,2, $J_{Fi}$ is the Fi layer exchange-field strength, and Ω is the precession frequency. We assume that the precession frequency is the same in both Fi layers. To model non-equilibrium and non-stationary properties in diffusive superconductors, we should use time-dependent non-equilibrium Usadel equations [12] which are derived from quasiclassical equations [8, 39]. The exchange field time dependency described above allows to transform non-stationary Usadel equations in the laboratory frame into stationary Usadel equations in the rotating frame [12]. Stationary non-equilibrium Usadel equations in the rotating frame write [12, 39]:

$$\frac{D}{\pi}\partial_x(\check{G}\,\partial_x\check{G}) + \left[E\hat{\tau}_3 - \frac{\Omega'}{2}\sigma^z - \check{\Sigma}, \check{G}\right] = 0 \quad (5)$$

where $D$ is the diffusion coefficient, $\sigma^z(\tau_3)$ the third Pauli matrix acting on the spin (particle-hole) subspace, $[,]$ is the commutator and $\check{G}$ and $\check{\Sigma}$ describes the normal and anomalous Green's function (self-energies) in the Keldysh space [8, 12, 39].

We rotate the axis such that the F1 layer magnetization points in the z-axis direction implying a reduced non-equilibrium exchange field $\Omega \to \Omega' = \Omega \cos(\theta_{F1})$ [12]. For notational simplicity, we do not explicitly state identity matrices. For inner interfaces, we assume (current conserving) Nazarov boundary conditions [39]:

$$\sigma_l \check{G}_l \partial_x \check{G}_l = \sigma_r \check{G}_r \partial_x \check{G}_r$$

$$\sigma_l \check{G}_l \partial_x \check{G}_l = \frac{1}{SR_b} \frac{2\pi^2 \tau [\check{G}_l, \check{G}_r]}{4\pi^2 - \tau(\{\check{G}_l, \check{G}_r\} + 2\pi^2)}$$

(6)

Where $G^{l(r)}$ is the Green's function on the left(right) side of the interface, $\sigma$ is the normal state electrical conductivity, $S$ is the area of the junction, $R_b$ is the interface resistivity and $0 < \tau < 1$ the interface transparency [39]. For the outer F1 interface, we impose the Green's function to be the one for a bulk ferromagnetic material with a spin-resolved non-equilibrium distribution function $f_{\uparrow(\downarrow)} = f_{FD}\left(E + (-)\frac{\Omega'}{2}\right)$ with $f_{FD}$ the Fermi Dirac distribution. At the outer S interface(x=L), we impose a vanishing-current boundary condition $\partial_x \check{G}|_{x=L} = 0$ with $L$ the trilayer thickness.

For the spatially varying superconducting order parameter we solve the self-consistency equation

$$\Delta(x) = \frac{\int_{-\infty}^{\infty} \frac{dE}{4i\pi} f_s^K(E, x)}{\int_{-\infty}^{\infty} \frac{dE}{2E} \tanh\left(\frac{E}{2T}\right) + \ln\left(\frac{T}{Tc}\right)}$$

(7)

where $f^K_s$ is the singlet part of the Keldysh anomalous Green function [8, 12, 39].

Green's functions are calculated by solving numerically the Usadel equations together with the boundary conditions and the superconducting self-consistency equation. The Green's functions are then used to calculate spin currents. Spin currents is obtained from the relation [12]:

$$I_s = I_s^0 \int_{-\infty}^{+\infty} dE \; \text{Tr}\left[\hat{\tau}_3 \boldsymbol{\sigma}(\check{G}\,\partial_x\check{G})^K\right] \tag{8}$$

with $I_s^0 = \frac{\hbar e N_0 D}{16\pi^2}$, $N_0$ the density of state at the Fermi energy, $e$ the electrical charge, $\hbar$ the reduced Planck constant, and σ=(σ$^X$,σ$^Y$,σ$^Z$) the spin Pauli matrix vector. The spin current vector is given by $I_s = (I_s^X, I_s^Y, I_s^Z)$ in terms of the spin Pauli matrix basis.

We solve the Usadel equation for a precessing F1 and F2 layers and we calculate the spin current $I_s^Z$ for all temperatures. Then, we calculate the Gilbert damping $\alpha_t$. The Gilbert damping $\alpha_t$ can be expressed in the form $\alpha_t = \alpha_0 + \alpha \equiv \alpha_0 + \beta I_s^Z$ where $\alpha_0$ describes an intrinsic Gilbert damping independent of temperature and superconducting properties, and $\alpha$ describes the additional Gilbert damping due to spin injection; the latter is proportional to the dissipative part of the spin current [42], which in itself is proportional to $I_s^Z$ in the F1 layer with a coefficient depending on the tip angle. The quantity $\alpha/\alpha_{T>T_c}^{\theta=0} = I_s^Z/I_{s,T>T_c}^{\theta=0,Z}$ is independent of the tip-angle dependent quantity β and therefore quantifies the additional Gilbert damping in our system.

**Acknowledgements**

This work was supported by the EPSRC Programme Grant EP/N017242/1, EPSRC Centre for Doctoral Training in Advanced characterisation of Materials Grant EP/L015277/1, and The Leverhulme Trust RPG-2016-306.

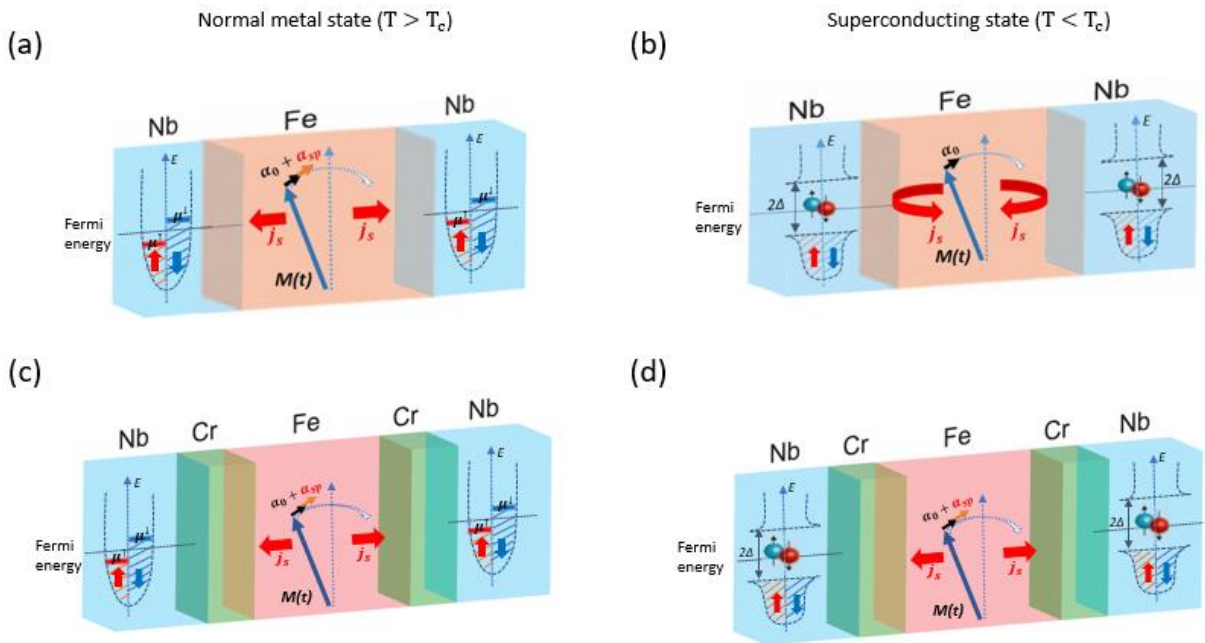

**Figure. 1.** Schematic diagrams of spin pumping in Nb/Fe/Nb and Nb/Cr/Fe/Cr/Nb heterostructures above (a,b) and below (c, d) the critical superconducting transition temperature, $T_c$, where (d) indicates the formation of triplets with the disorder Fe/Cr spin interface and (c) the lack of triplet generation for the ordered Fe/Nb interface. Spin-dependent density of states and its occupation in SCs are indicated by the red (majority-spin) and blue (minority- spin) symbols. $M(t)$, $J_s$ and $\alpha_{sp}(\alpha_0)$ represent the time-varying magnetization vector of the FM, the spin current injected from the FM into the SC by spin pumping, and the Gilbert damping of the FM relevant (irrelevant) to the spin pumping, respectively. $\mu^\uparrow$, $\mu^\downarrow$ and $\Delta$ are the spin dependent electrochemical potentials and superconducting energy gap, respectively.

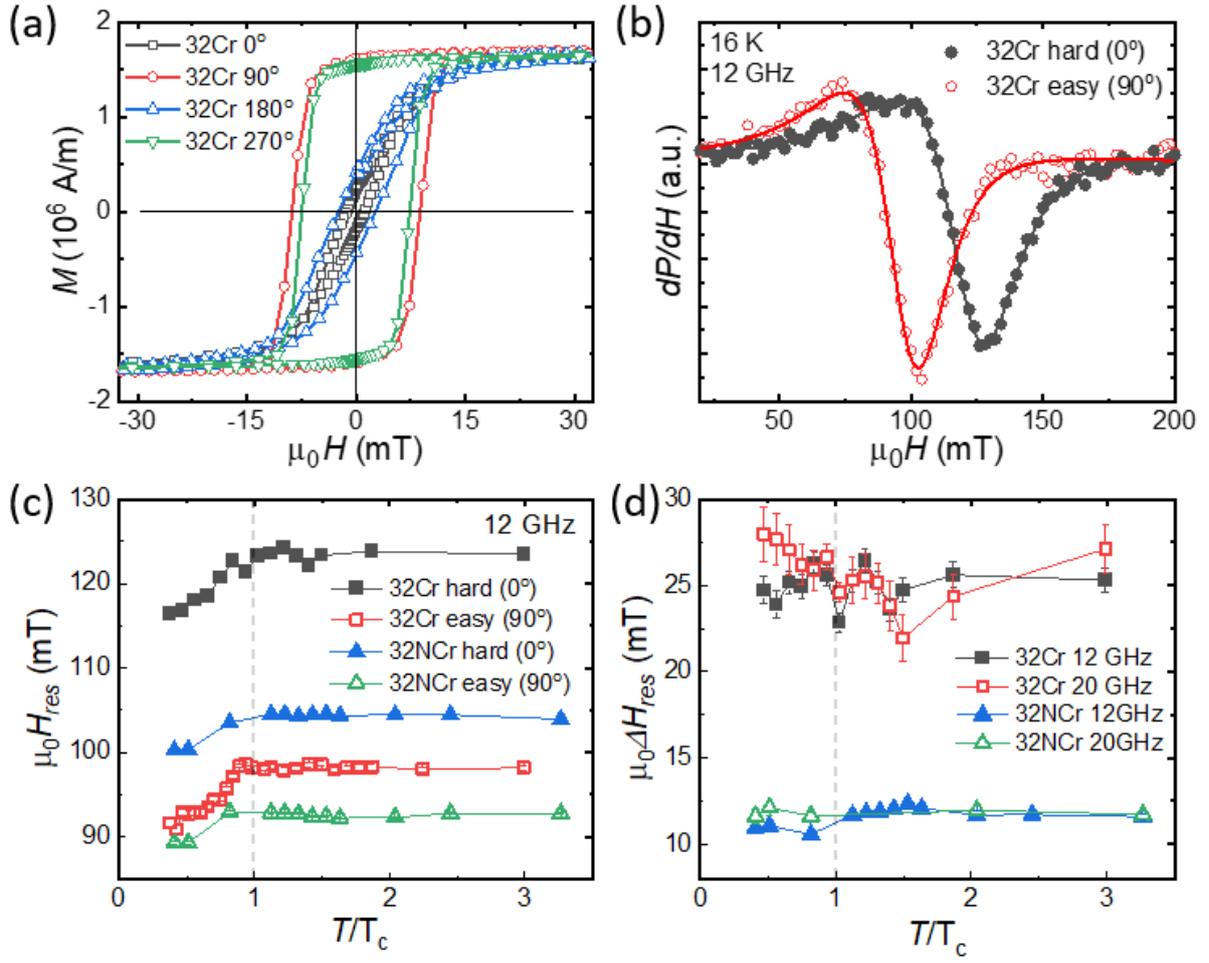

**Figure 2.** (a) M(H) curves for film 32b in four orientations with respect to the applied in-plane field measured at 300 K. (b) FMR absorption power derivative $(dP/dH)$ for sample 32Cr measured at 16 K. (c) $\mu_0 H_{res}$ as a function of temperature for film 32Cr and 32NCr extracted from FMR measurements. (d) The temperature dependence of the difference in $\mu_0 H_{res}$ between the hard and easy orientations ($\mu_0 \Delta H_{res}$) for films 32Cr and 32NCr.

| Sample Structure | Ref | Easy axis $\mu_0 H_c$ (mT) | Hard axis $\mu_0 H_c$ (mT) | Easy axis $M_r/M_s$ | Hard axis $M_r/M_s$ | Ms ($10^3$ A/m) | $K_{VSM}$ (J/m$^3$) | $K_{FMR}$ (J/m$^3$) |
|---|---|---|---|---|---|---|---|---|
| Nb(32)/Cr(1)/Fe(6)/Cr(1)/Nb(32) | 32Cr | 7.4 | 1.1 | 0.93 | 0.19 | 1680 | 7807 | 8193 |
| Nb(32)/Fe(6)/Nb(32) | 32NCr | 8.3 | 3.4 | 0.90 | 0.50 | 1500 | 7438 | 4880 |
| Nb(31)/Cr(1)/Fe(6)/Cr(1)/Nb(31) | 31Cr | 9.8 | 6.0 | 0.95 | 0.68 | 1480 | 5153 | 5129 |
| Nb(31)/Fe(6)/Nb(31) | 31NCr | 9.4 | 9.0 | 0.97 | 0.79 | 1250 | 1348 | 2763 |

**Table 1**. Samples and their magnetic property values. $K_{VSM}$ and $K_{FMR}$ are the anisotropy constant values obtained from the VSM MH loop and the FMR measurements respectively.

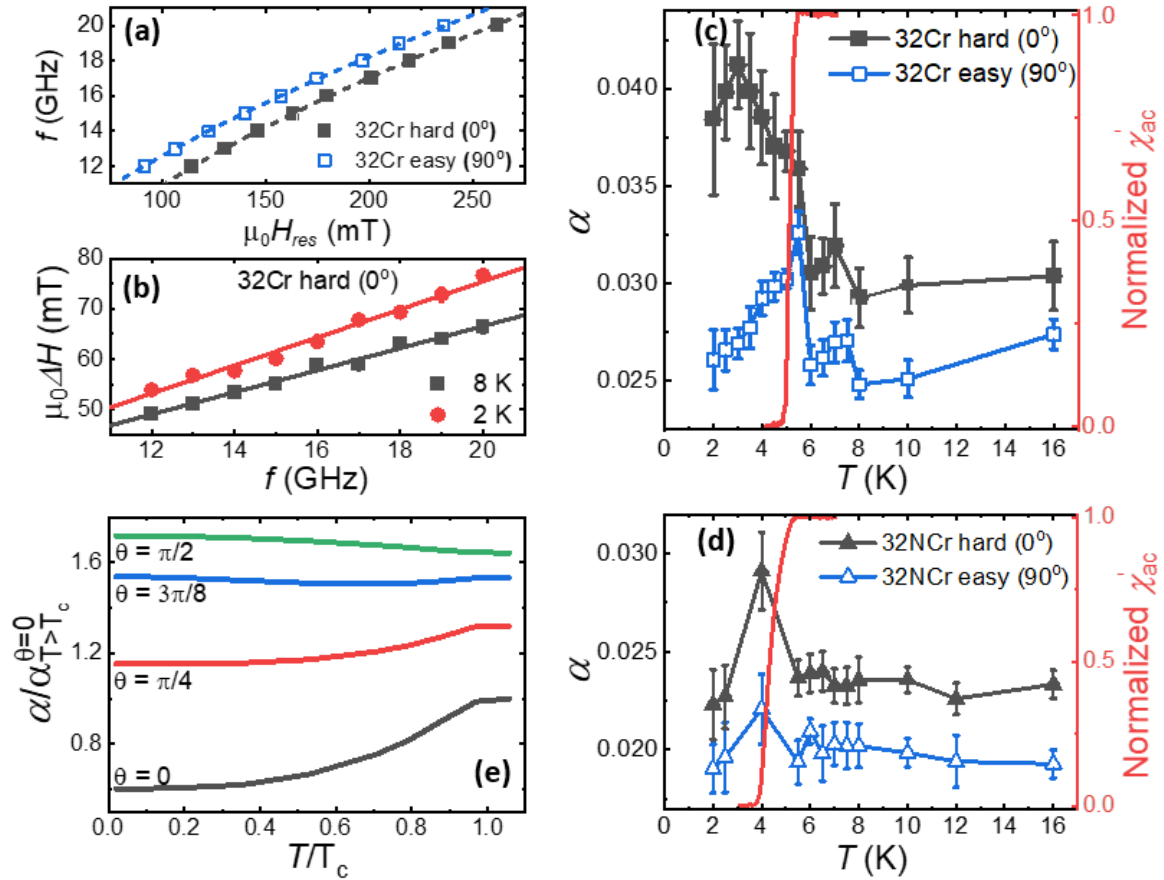

**Figure 3.** (a) Microwave frequency dependence of the resonance field $\mu_0 H_{res}$ for sample 32Cr, with the field applied in the hard axis orientation (black filled squares) and the easy axis orientation (blue open squares), fitted to a modified Kittel fit (dashed lines) including the anisotropy $H_2$ and superconducting shift term $\mu_0 H_{shift}^{SC}$. (b) Microwave frequency dependence of the linewidth $\mu_0 \Delta H$ for sample 32b with the field applied in the hard axis orientation, above (8 K in black) and below (2 K in blue) $T_c$. (c) Temperature dependence of the Gilbert damping $\alpha$ for sample 32Cr in the hard and easy orientation. (d) Temperature dependence of $\alpha$ for sample 32NCr in the hard and easy orientation. (e) Theoretical calculation of the temperature dependence of the Gilbert damping parameter $\alpha/\alpha_{T>T_c}$ for various F2 layer magnetization tilting angles $\theta_{F2} = 0$ (black), $\theta_{F2} = \pi/4$ (red), $\theta_{F2} = 3p/8$ (blue) and $\theta_{F2} = \pi/2$ (green).

# Supplementary information (SI)

## Section S1. Magnetization measurements on Nb/Cr/Fe/Cr/Nb and Nb/Fe/Nb samples at room temperature.

Magnetic field loops (M-H loops) of the samples were measured with an external dc magnetic field ($H_{ext}$) applied in-plane to the samples at room temperature (300 K). To investigate the anisotropy of the samples, $H_{ext}$ is applied in four orientations 90º to each other. M-H loops are shown in Fig. S1. All four samples show uniaxial anisotropy to varying strengths.

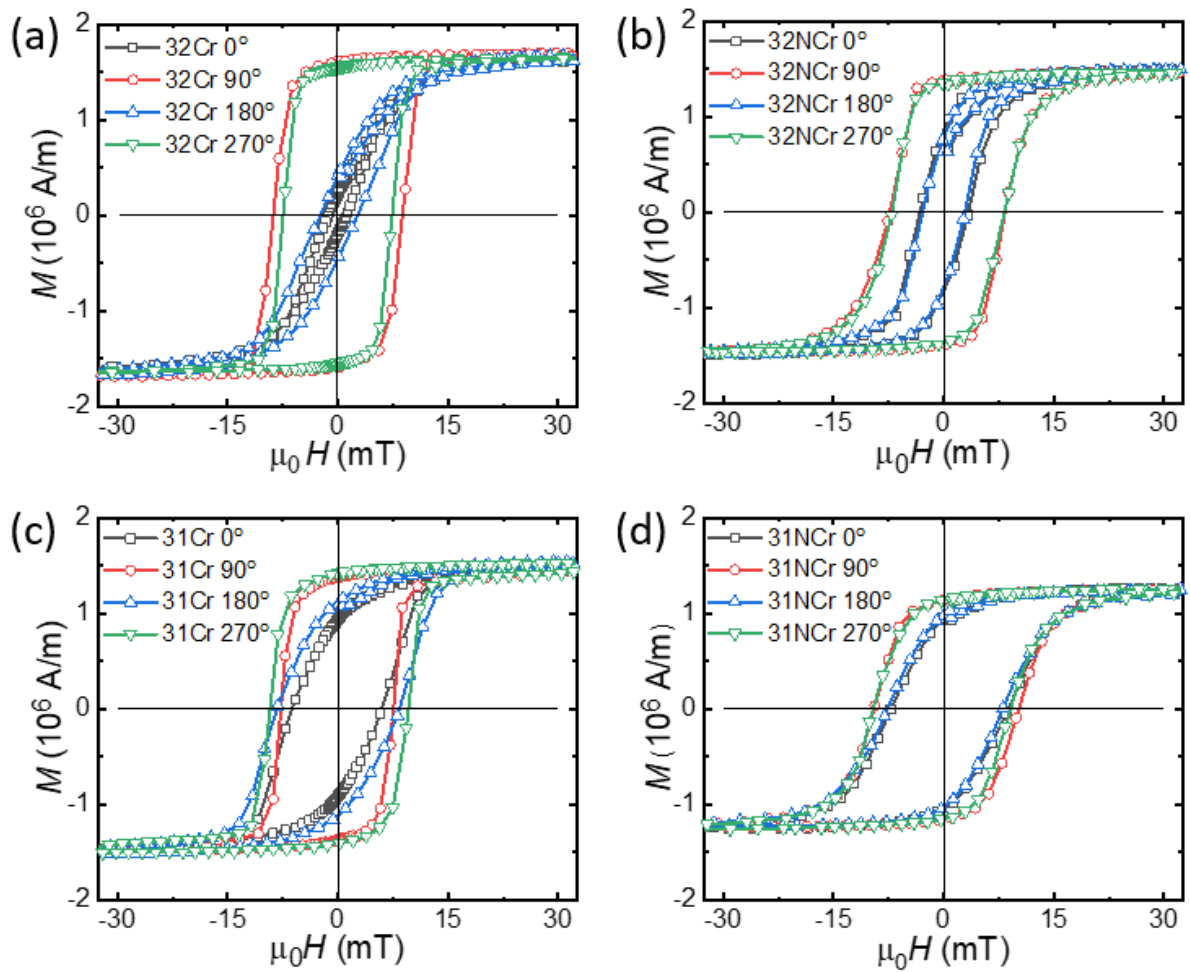

**Figure S1 Magnetic field loops** Magnetic field loops (M-H loops) in four orientations with respect to the applied in-plane field measured at 300 K for films (a) 32Cr (b) 32NCr (c) 31Cr and (d) 31NCr.

## Section S2. Superconducting transition temperature $T_c$ for Nb/Cr/Fe/Cr/Nb and Nb/Fe/Nb structures.

The full set of critical temperature ($T_c$) curves for each sample structure are presented in Fig. S2, measured by surface contact resistivity in Fig. S2(a), and by bulk ac magnetic susceptibility with no dc field in Fig. S2(b). In-phase susceptibility ($\chi'_{ac}$) is shown. To show that both measurements are in good agreement, they are compared in Fig. S2(c) with the resistivity temperature offset by -1 K due to a difference in thermometer calibration.

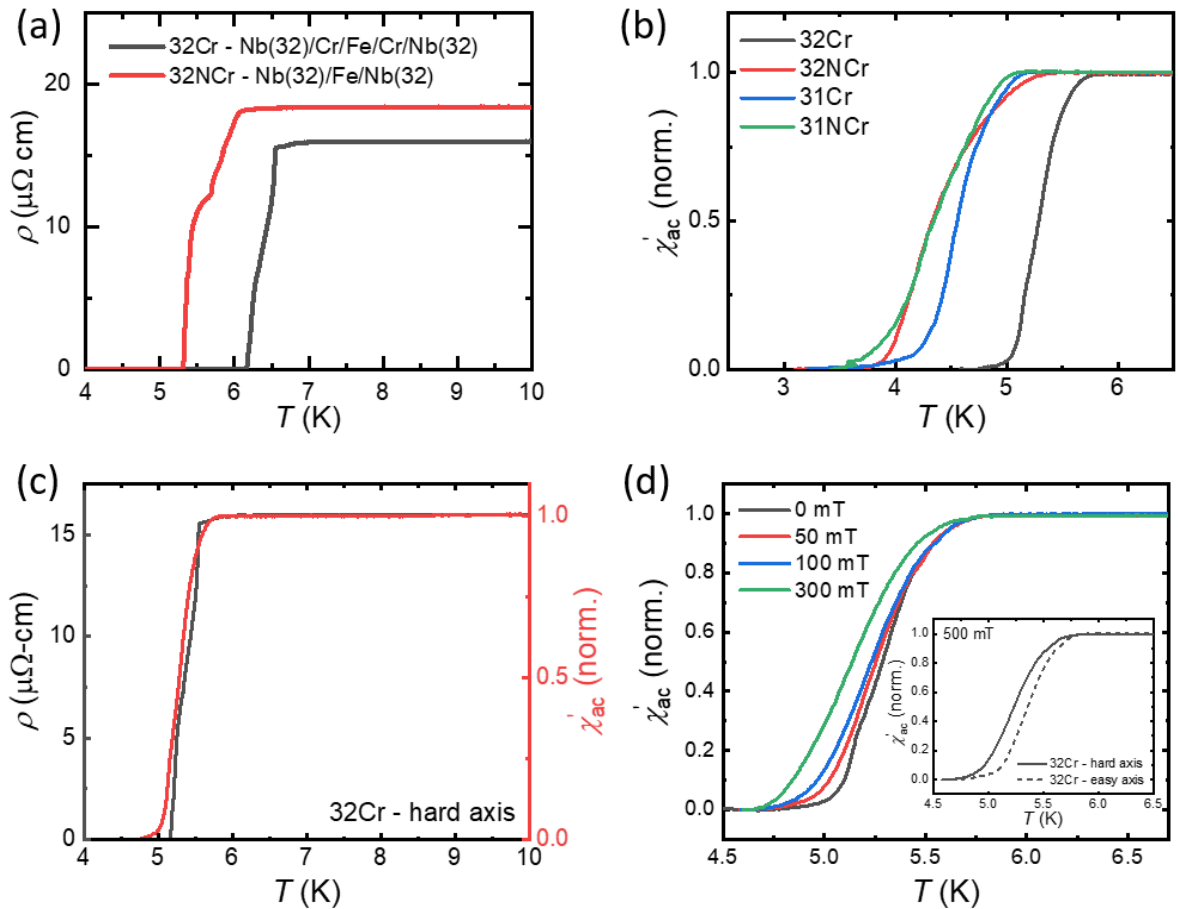

**Figure S2 Critical temperature measurements of superconducting samples** (a) The temperature ($T$) dependence of the resistivity ($\rho$) for 32Cr (black) and 32NCr (red). (b) $T$ dependence of the normalised in-phase component of the ac magnetic susceptibility ($\chi'_{ac}$) for all four samples. (c) Comparison of the $T$ dependence of $\rho$ and $\chi'_{ac}$ for 32Cr. (d) $T$ dependence of $\chi'_{ac}$ with various dc fields applied in-plane along the hard axis orientation for 32Cr. The inset shows $\chi'_{ac}(T)$ for 32Cr with a 500 mT field applied in-plane along the hard (solid line) and easy (dashed line) orientations.

The ac magnetic susceptibility measurements were performed in the same QuantumDesign Physical Property Measurement System® (PPMS®) as ferromagnetic resonance (FMR) measurements, and so we take the $T_c$ value from those measurements.

To characterize the $T_c$ transition of these samples in external in-plane dc fields of typical magnitudes applied in FMR measurements, ac susceptibility was measured at various $H_{ext}$ of 0 mT, 50 mT, 100 mT, and 300 mT. These measurements are shown in Fig. S2(d) for 32Cr. The application of $H_{ext}$ up to 300 mT cause minimal change in the $T_c$ transition, only broadening in field by a maximum of 0.2 K. The $T_c$ transition is also independent of the field orientation, as shown in the inset of Fig. S2(d). In both the easy and hard axis case for each sample, $\chi'_{ac}$ begins to decrease at the same $T$ indicating the onset of superconductivity. From the $T$ derivative of $\chi'_{ac}$, $T_c$ is taken to be the value of $T$ that gives a maximum slope.

In both the easy and hard axis case, the Gilbert damping $\alpha$ is constant at temperatures where the normalised $\chi'_{ac}$ value is 1. Only when $\chi'_{ac}$ starts to decrease, indicating the onset of superconductivity, does $\alpha(T)$ increase. This provides further evidence that the superconducting Nb affects the magnetisation dynamics of the system. In the easy axis case, $\alpha(T)$ increases sharply around the same temperature that $\chi'_{ac}$ decreases. $\alpha(T)$ then decreases within the $T_c$ transition width. This reflects the picture of enhanced availability of quasiparticle states as the superconducting gap opens [S1], but gradually decreases with $T$ as the quasiparticle states freeze out due to the superconducting gap opening further [S1, S2]. In the hard axis case, $\alpha(T)$ begins to increase at the same temperature as $\chi'_{ac}$ decreases and continues to increase beyond the $T_c$ transition width. This shows that $\alpha(T)$ increases below $T_c$ and well after the opening of a superconducting gap. Therefore, it cannot be explained by the quasiparticle spin transport, but can be attributed to long-range proximity induced spin-triplets.

**Section S3. Microwave frequency dependence of FMR spectra for Nb/Cr/Fe/Cr/Nb and Nb/Fe/Nb structures**

Fig. S3 displays typical FMR spectra attained from the samples with Nb/Cr/Fe/Cr/Nb and Nb/Fe/Nb at 16 K from which the data shown in the Fig. 3 were extracted. To accurately determine the resonance magnetic field $\mu_0 H_{res}$ and the FMR (peak-to-peak) linewidth $\mu_0 \Delta H$, we fitted all FMR spectra with a Lorentzian derivative:

$$\frac{dP}{dH} = -V_{sym} h_{ac} \frac{2(H_{ext} - H_{res})\Delta H^2}{[\Delta H^2 + (H_{ext} - H_{res})^2]^2} - V_{asym} h_{ac} \frac{\Delta H((H_{ext} - H_{res})^2 - \Delta H^2)}{[\Delta H^2 + (H_{ext} - H_{res})^2]^2}, \quad (S1)$$

Where $V_{sym}$ and $V_{asym}$ denote the symmetric and antisymmetric Lorentzian components, $h_{ac}$ is the small modulation field provided by the Helmholtz coils.

The temperature dependence of $\mu_0 H_{res}$ for samples 31Cr and 31NCr are shown in Fig. S4(a), and the difference in resonance fields $\mu_0 \Delta H_{res}$ between when $H_{ext}$ lies along easy and hard directions is shown in Fig. S4(b).

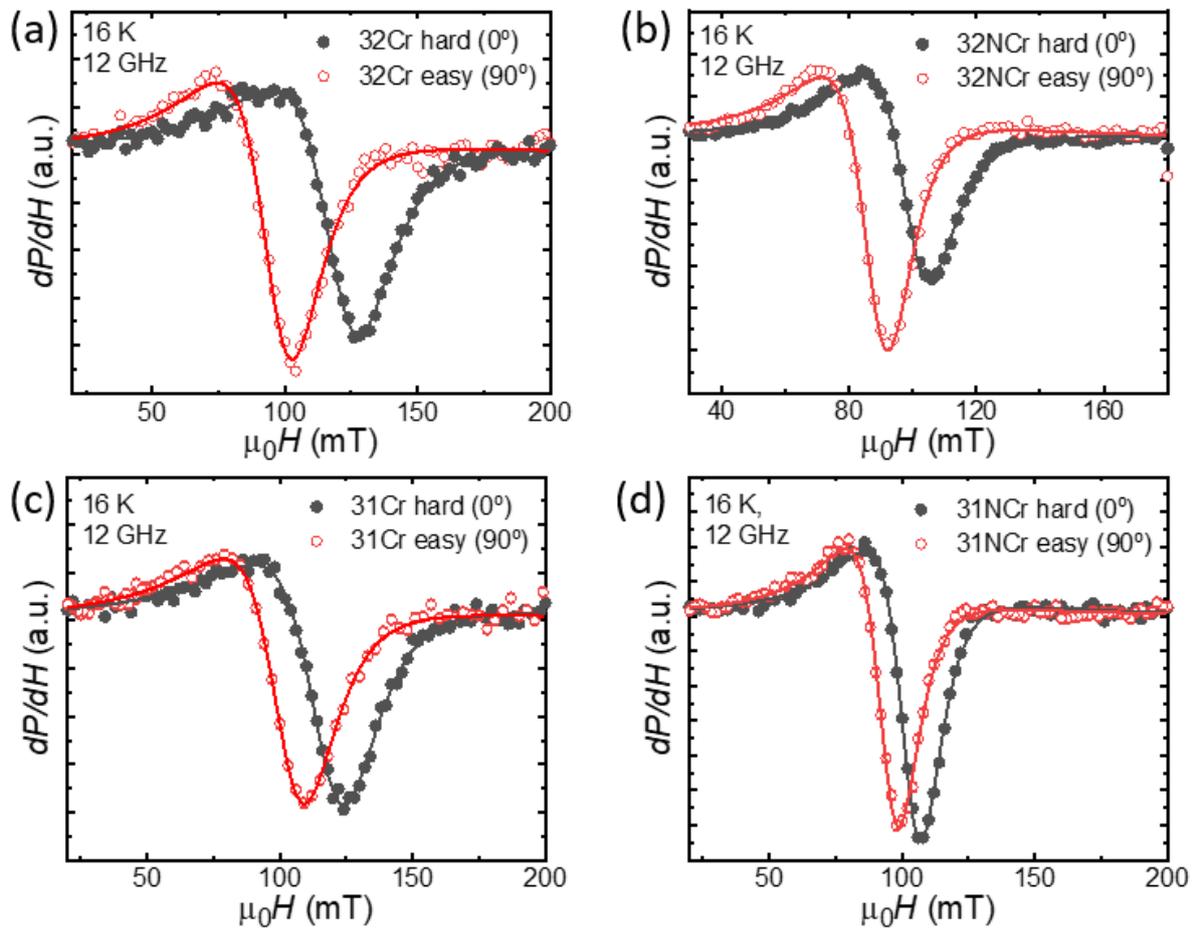

**Figure S3 Ferromagnetic resonance (FMR) spectra** FMR absorption power derivative (*P*) measured at 16 K and 12 GHz for samples (a) 32Cr (b) 32NCr (c) 31Cr and (d) 31NCr.

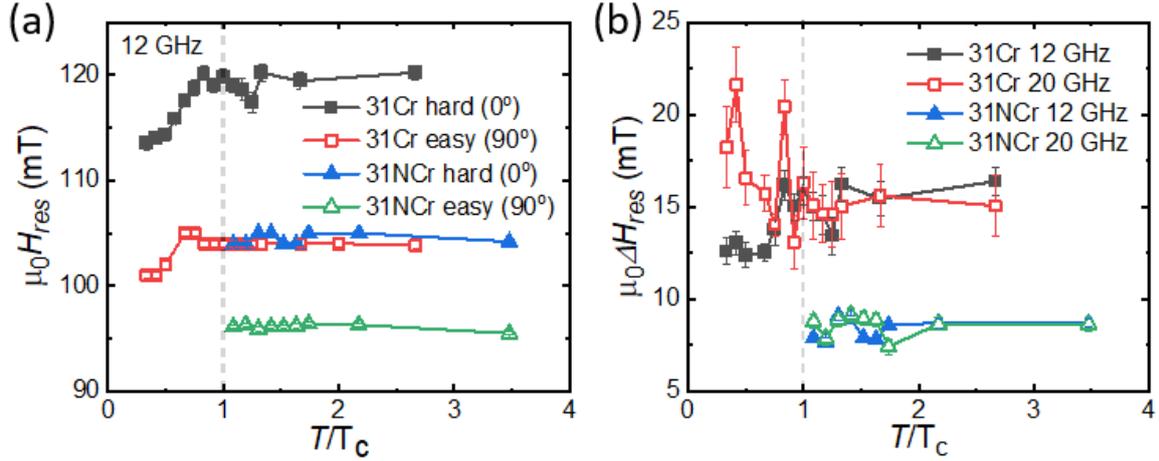

**Figure S4 Temperature dependence of the resonance field** (a) $\mu_0 H_{res}$ as a function of temperature for film 32Cr and 32NCr extracted from FMR measurements. (b) The temperature dependence of the difference in $\mu_0 H_{res}$ between the hard and easy orientations $\mu_0 \Delta H_{res}$ for films 32Cr and 32NCr.

Similar to samples 32Cr and 32NCr (Fig. 2(c) and 2(d)), the samples with Cr demonstrate a higher $\mu_0 H_{res}$ and $\mu_0 \Delta H_{res}$. This further confirms the spin misorientation due to the presence of Cr next to Fe that survives up to $\mu_0 H_{res}$.

**Section S4. Frequency dependence of the linewidth**

The microwave frequency $f$ dependence of the linewidth $\mu_0 \Delta H$ extracted from the FMR spectra, at temperatures above $T_c$ and below $T_c$, for samples 32NCr and 31Cr are shown in figure S5. These are fitted to Equation 2 (shown in the main text) and show the clear greater gradient ($\alpha$) for samples with Cr when cooled below the samples' $T_c$. $\alpha$ is extracted as the gradient of these fits and plotted as a function of $T$ for samples 31Cr and 31NCr in figure S6. Similar to the 32 series samples, the sample with Cr has a higher average $\alpha$ and only the sample with Cr shows an apparent rise in $\alpha$ below the sample's $T_c$. The spin misorientation layer provides an additional spin relaxation channel, leading to the observed $\alpha(T)$ relationship. Further evidence of the spin misorientation layer is given by the magnetic inhomogeneity ($\mu_0 \Delta H_0$), extracted from Equation 2 and are shown for all samples as a function of T in Fig. S7. The samples containing Cr demonstrate a higher $\mu_0 \Delta H_0$. 32Cr and 31Cr both have $\mu_0 \Delta H_0 \sim 22$ mT while 32NCr and 31NCr have $\mu_0 \Delta H_0 \sim 12-15$ mT. This indicates that the addition of Cr creates a spin misorientation layer and therefore increases $\mu_0 \Delta H_0$. Only the samples containing Cr (32Cr and 31Cr) show the possibility of an increased $\alpha$ below $T_c$ due to the additional spin relaxation channel.

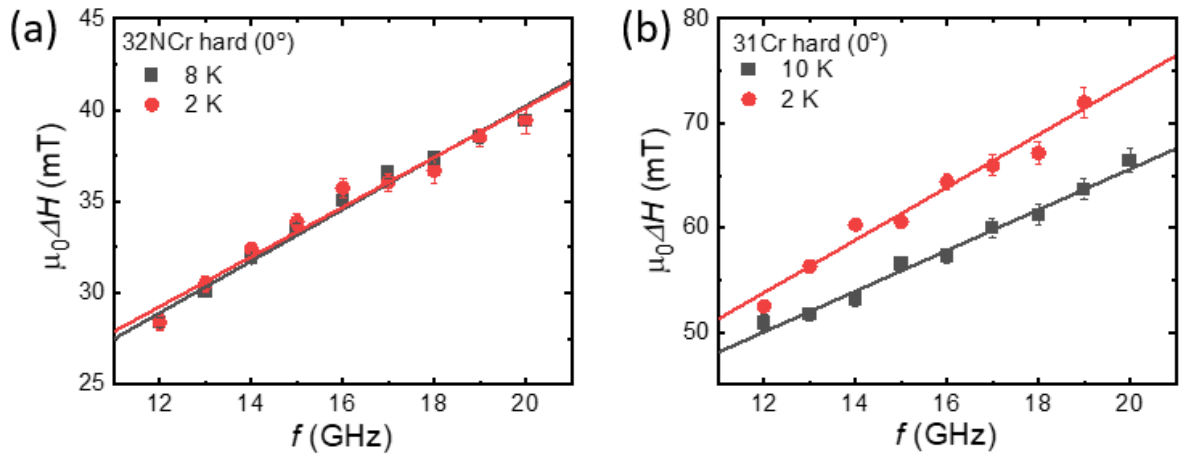

**Figure S5 Frequency dependence of the linewidth** Microwave frequency dependence of the linewidth (μ$_0$ΔH) for sample (a) 32NCr and (b) 31Cr with the field applied in the hard axis orientation, above (8 K in black) and below (2 K in red) $T_c$. Solid lines are fits to equation 2.

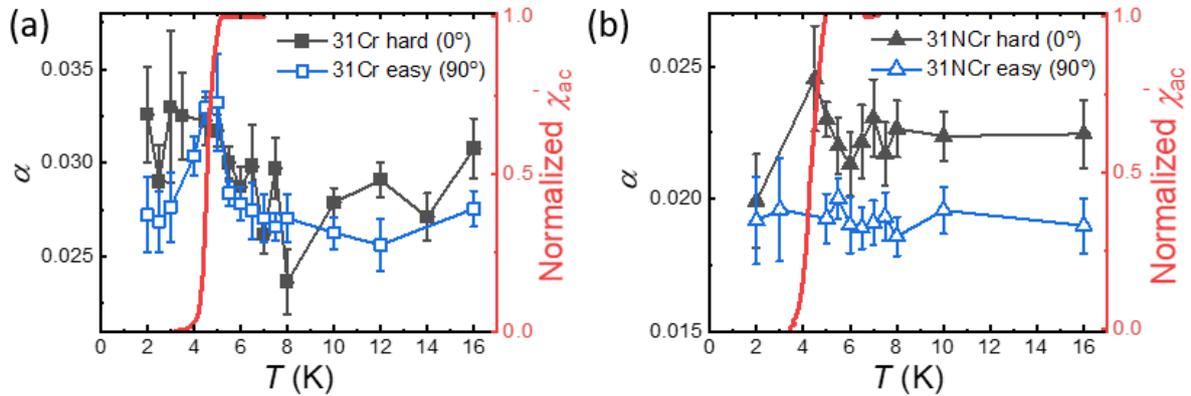

**Figure S6 Gilbert damping of samples 31Cr and 31NCr** Temperature dependence of the Gilbert damping ($\alpha$) for (a) 31Cr and (b) 31NCr, with $\chi'_{ac}$ in red.

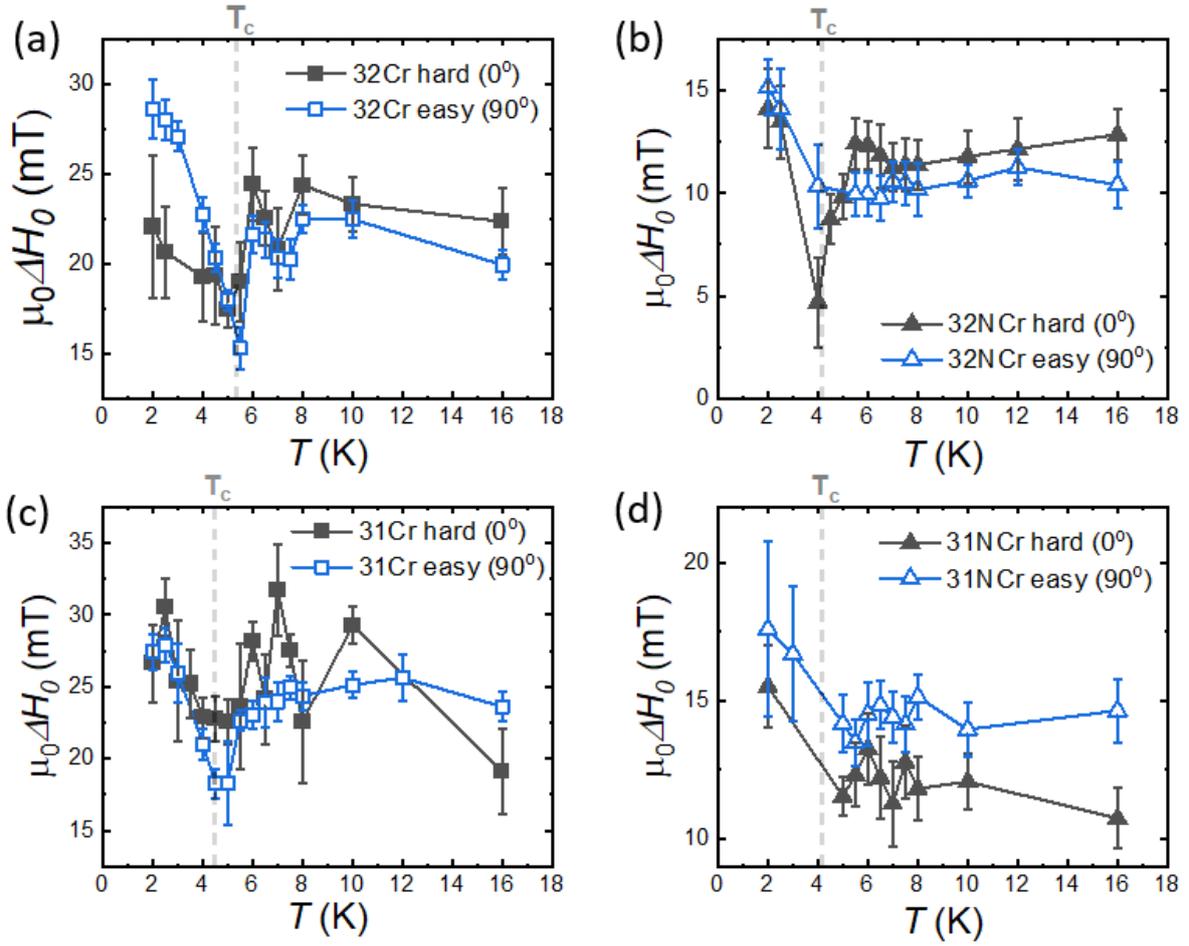

**Figure S7 Magnetic inhomogeneity for the four samples** Temperature dependence of the magnetic inhomogeneity (μ$_0$ΔH$_0$) for samples (a) 32Cr, (b) 32NCr, (c) 31Cr, (d) 31NCr with the external magnetic field applied in the hard (black) and easy (blue) directions.

**Section S5. Frequency dependence of the resonance field $\mu_0 H_{res}$**

The *f* dependence of μ$_0$H$_{res}$ can be fitted to the modified Kittel formula, taking into consideration the magnetic anisotropy field contribution [S3]:

$$f^2 = \left(\frac{\gamma \mu_0}{2\pi}\right)^2 [\{M_{\text{eff}} + H_{\text{res}} + H_k \sin^2(\phi - \phi_0)\} \times \{H_{\text{res}} - H_k \cos 2(\phi - \phi_0)\}], \quad (S2)$$

where γ is the gyromagnetic ratio, *M*$_{eff}$ is the effective saturation magnetisation, ϕ is the angle of the film's magnetisation with respect to the easy axis direction, ϕ$_0$ is the angle of the hard axis direction with respect to the easy axis direction, and all other symbols take their usual definitions. In the case of the superconducting state ($T < T_c$), it may be necessary to include a superconducting shift term $\mu_0 H^{SC}_{shift}$ [S4]:

$$f^2 = \left(\frac{\gamma \mu_0}{2\pi}\right)^2 [\{M_{\text{eff}} + H_{\text{res}} + H_k \sin^2(\phi - \phi_0) + H^{SC}_{\text{shift}}\} \\ \times \{H_{\text{res}} - H_k \cos2(\phi - \phi_0) + H^{SC}_{\text{shift}}\}] \tag{S3}$$

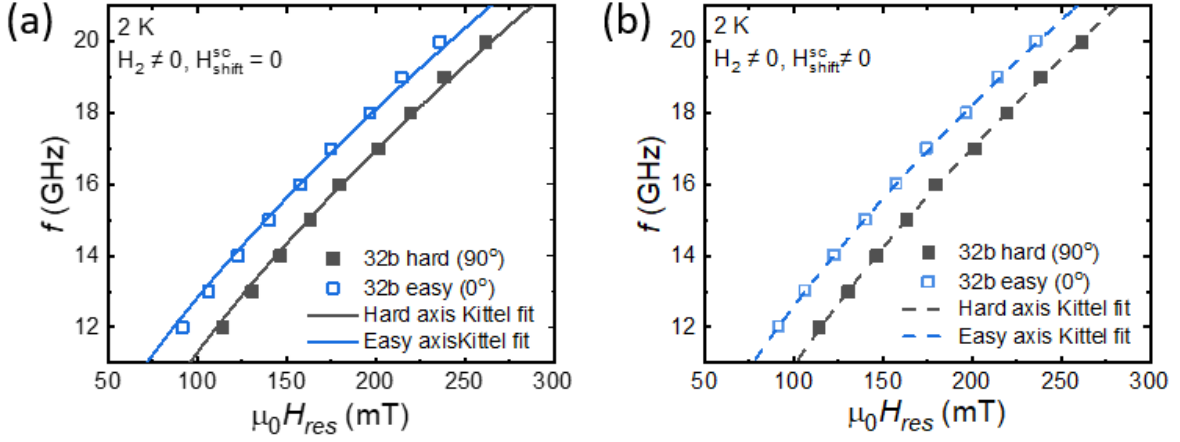

**Figure S8 Resonance field at different frequencies** Microwave frequency dependence of the resonance field ($\mu_0 H_{res}$) for sample 32Cr, fitted to the Kittel formula (a) without and (b) with the $\mu_0 H^{SC}_{shift}$ term.

This term arises from the onset of superconductivity, the Meissner effect, and vortices of pinned flux contributing to internal dc magnetic fields [S4]. Flux pinning forms in type II superconductors, such as Nb, when an external magnetic field is applied and trapped as quantized flux pins within the SC layer. The magnetic field strength must be between the superconductor's lower and upper critical field ($H_{c1} < H_{ext} < H_{c2}$). The radius of each vortex is equivalent to the London magnetic penetration depth ($\lambda_L$). Therefore, we would not expect flux pinning to form in our samples with *H*$_{ext}$ applied in-plane where the Nb layers are thinner than the penetration depth of Nb ($\lambda_{Nb}$ = 39 nm) [S5]. However, we observe anomalous $M_{eff}(T)$ behaviour that can be explained by the addition of a $H^{SC}_{shift}$ term.

The frequency dependence of μ$_0 H_{res}$ for sample 32Cr is shown in Fig. S8, fitted without $\mu_0 H^{SC}_{shift}$ in Fig. S8(a), and fitted with $\mu_0 H^{SC}_{shift}$ in Fig. S8(b). The values of μ$_0 M_{eff}$ determined from Fig. S8 using Equations S2 and S3 for each sample between

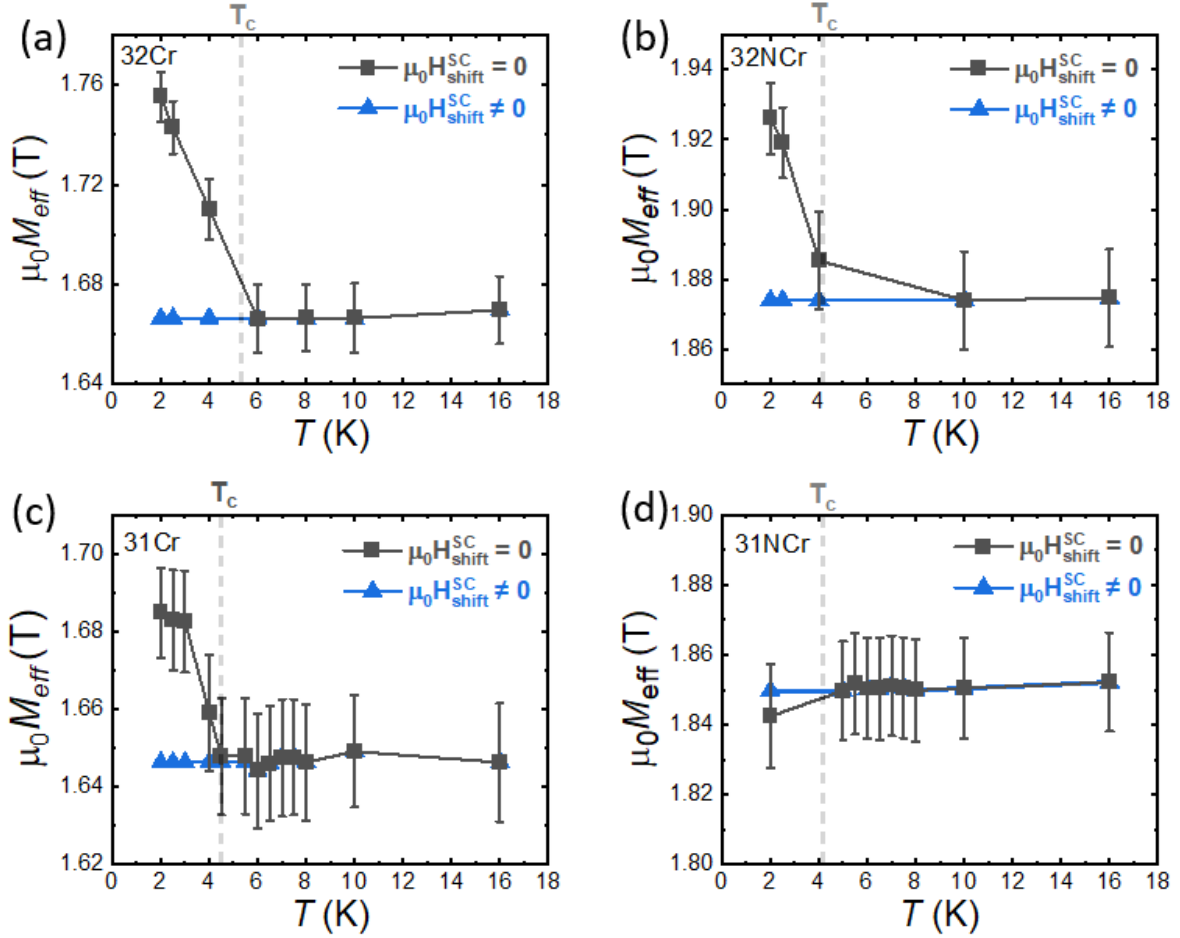

**Figure S9 Effective magnetisation for the four samples** Effective saturation magnetization extracted from Equation S3 as a function of temperature ($\mu_0 M_{eff}(T)$) for various samples from 16 K to 2 K for (a) 32Cr (b) 32NCr (c) 31Cr and (d) 31NCr, with (blue triangles) and without (black squares) $\mu_0 H_{shift}^{SC}$.

16 K and 2K are summarized in Fig. S9. The extracted values of $\mu_0 M_{eff}$ show an anomalous increase below $T_c$ when fitted without $\mu_0 H_{shift}^{SC}$, which is corrected when $\mu_0 H_{shift}^{SC}$ is considered. This is similar to [S4] and demonstrates the need for the $\mu_0 H_{shift}^{SC}$ term even in our samples of considerably thinner layers of superconducting Nb, well below the London penetration depth of Nb. This indicates that the applied external field is not perfectly in-plane with the sample, but that there is a small out-of-plane component which can be trapped as quantized flux.

**Section S6. Measuring different regions of the samples**

To ensure reproducibility and homogeneity on a mm length scale, the samples are measured in FMR across different regions with the microwave strip line placed in three different locations, labelled as top, middle, and bottom in Fig. S10. $\alpha$ and $\mu_0\Delta H_0$ for each region has been extracted and examples for sample 32Cr are shown in Fig. S11. It can be seen that the temperature dependence of $\alpha$ and $\mu_0\Delta H_0$ show consistent trends within experimental errors, indicating reproducibility and homogeneity in the samples (on a mm scale). $\alpha$ presented in the main text (Fig. 3) are taken from the averages of the linewidths at different sample regions.

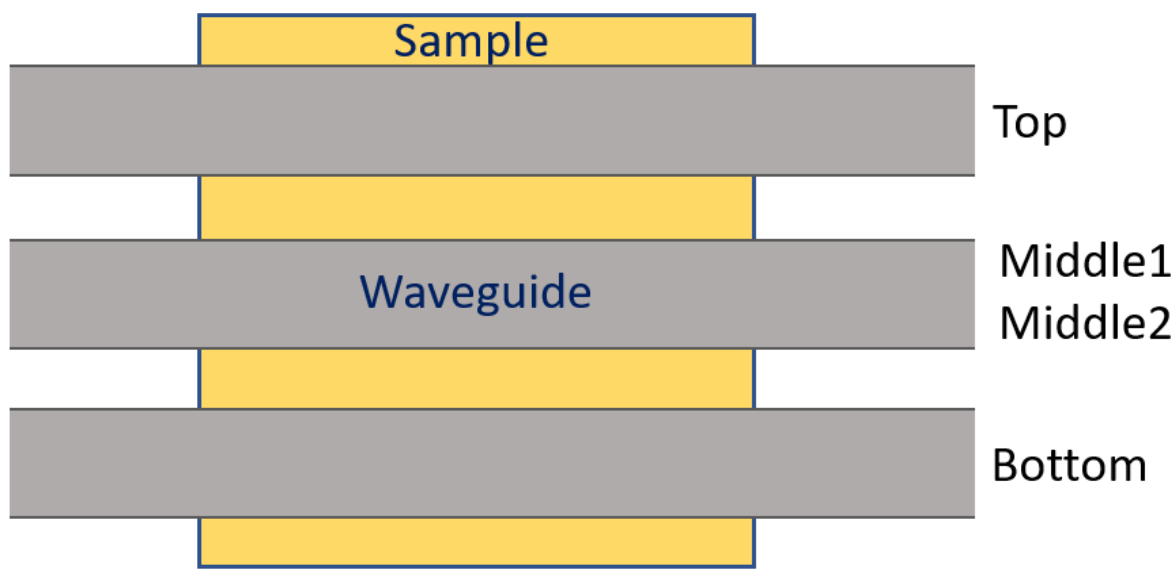

**Figure S10** Schematic of the three waveguide positions to investigate sample homogeneity in the mm scale.

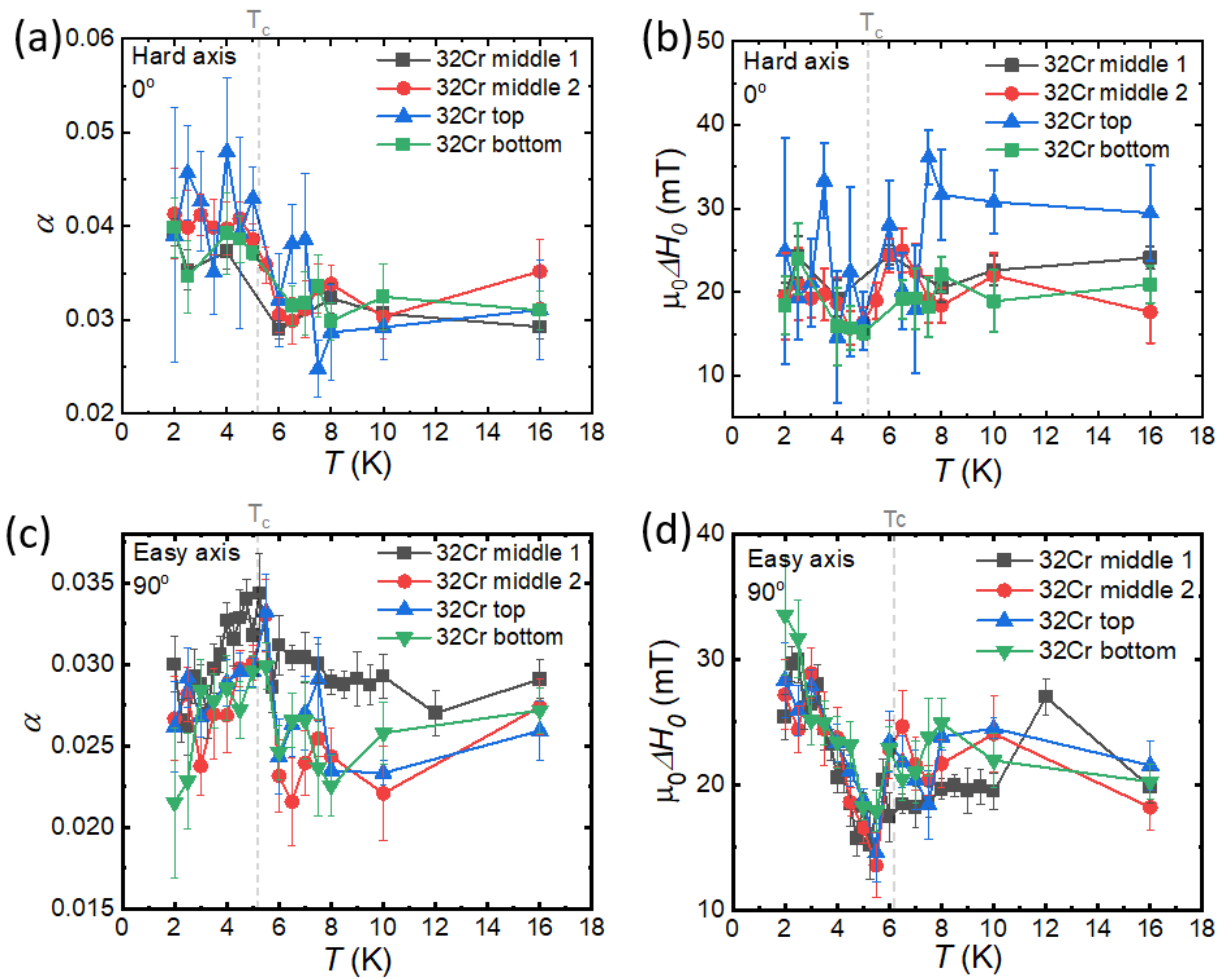

**Figure S11** Extracted values of $\alpha$ and $\mu_0\Delta H_0$ measured at different regions of the sample 32Cr. (a), (b), Temperature dependence of $\alpha$, $\mu_0\Delta H_0$, respectively, with the applied external field aligned with the sample hard axis. (c), (d), Temperature dependence of $\alpha$, $\mu_0\Delta H_0$, respectively, with the applied external field aligned with the sample easy axis.